\documentclass{ws-p8-50x6-00}

\begin{document}

\title{Summary of session on jets and inclusive hadron production}

\author{M. Wing}

\address{University of Bristol, DESY, Notkestrasse 85, 22607 Hamburg, Germany\\ 
E-mail: wing@mail.desy.de}

\maketitle

\abstracts{
Some highlights of the session on jets and hadron production are summarised. Results
between different experiments and measurements are compared and, in particular, how 
LEP and HERA complement each other discussed. Improvements in both theory and experiment  
for jet and hadron production and how the measurements can be used to extract 
fundamental parameters are addressed.
}

\section{Introduction}

Measurements of jets and hadron production at both LEP and HERA provide detailed 
investigations of many facets of Quantum Chromodynamics (QCD). They provide constraints 
on the parton densities in both the photon (LEP and HERA) and proton (HERA) in 
a complementary way to the inclusive measurements of the structure functions, $F_2^\gamma$ 
and $F_2^{\rm p}$. The cross sections are also sensitive to the dynamics of the 
hard sub-processes and perturbative QCD (pQCD) calculations. The transition of partons 
into hadrons can also be measured, thereby bridging the gap between 
pQCD calculations in terms of partons and the hadrons seen in detectors.

A lot of precise data exist in jet and hadron production from LEP and HERA, some of which 
are here discussed. For some results all of the available data from LEP has been analysed 
and at HERA the interpretation of many measurements is becoming limited by large 
theoretical uncertainties. To ensure that the data has a legacy, global fits to parton 
densities and Monte Carlo (MC) tuning need to be done incorporating much of this data. 
This will improve future measurements at both HERA and other colliders, in particular 
the LHC where very precise knowledge of the proton is needed.

\section{Summary of results}

\subsection{Hadron production}

\begin{figure}[htp]
\begin{center}
~\epsfig{file=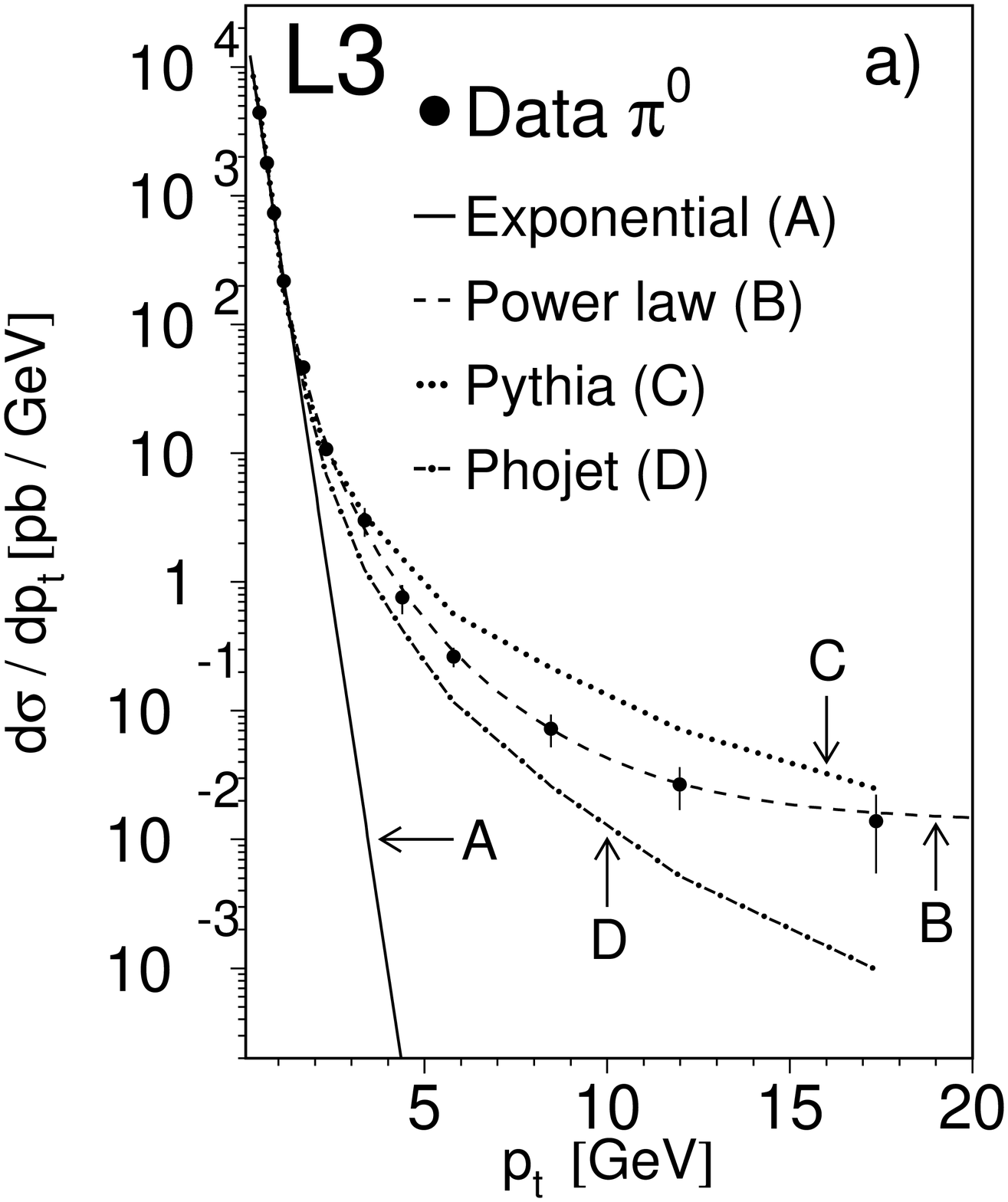,height=4.5cm}~\epsfig{file=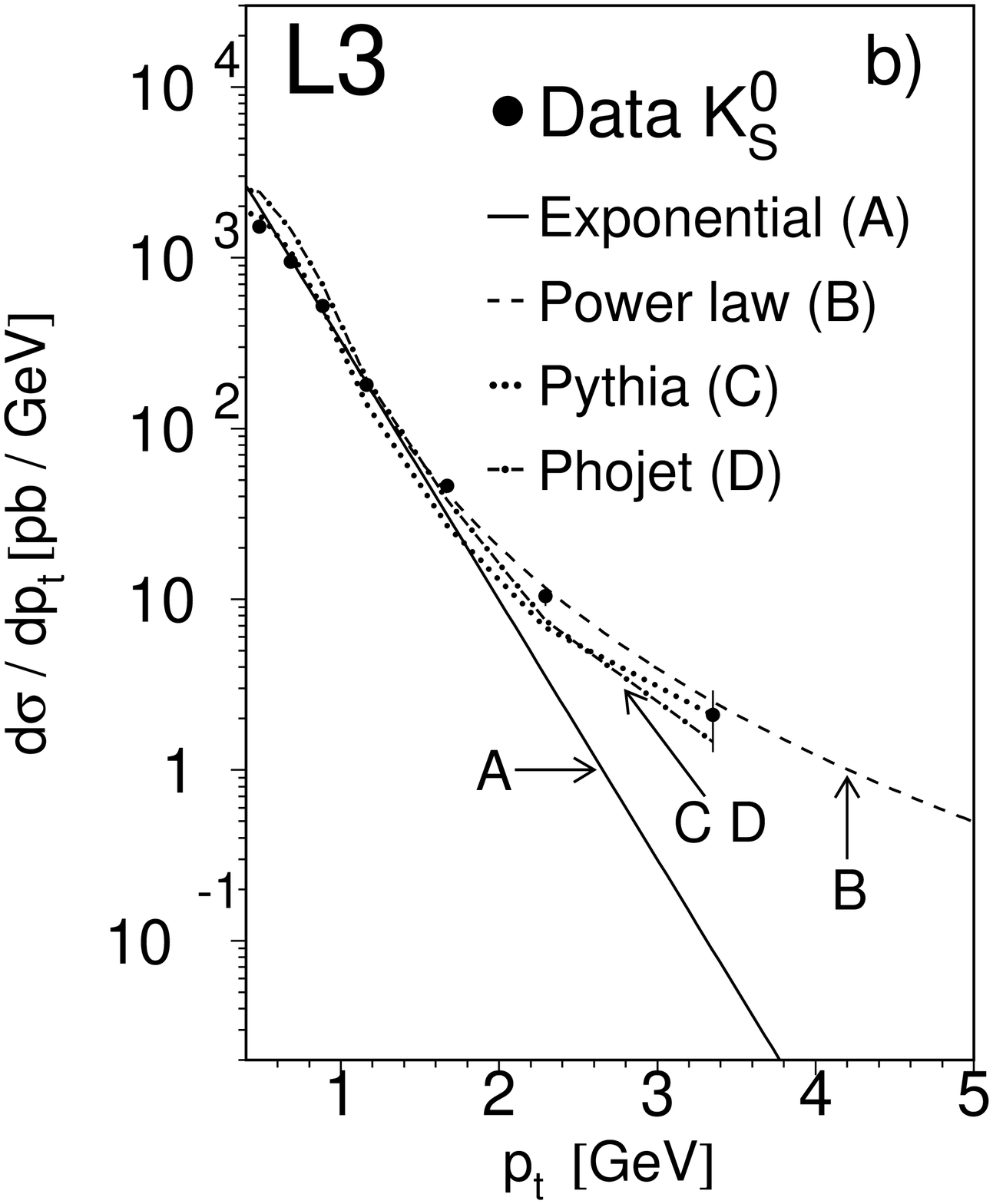,height=4.5cm}
~\epsfig{file=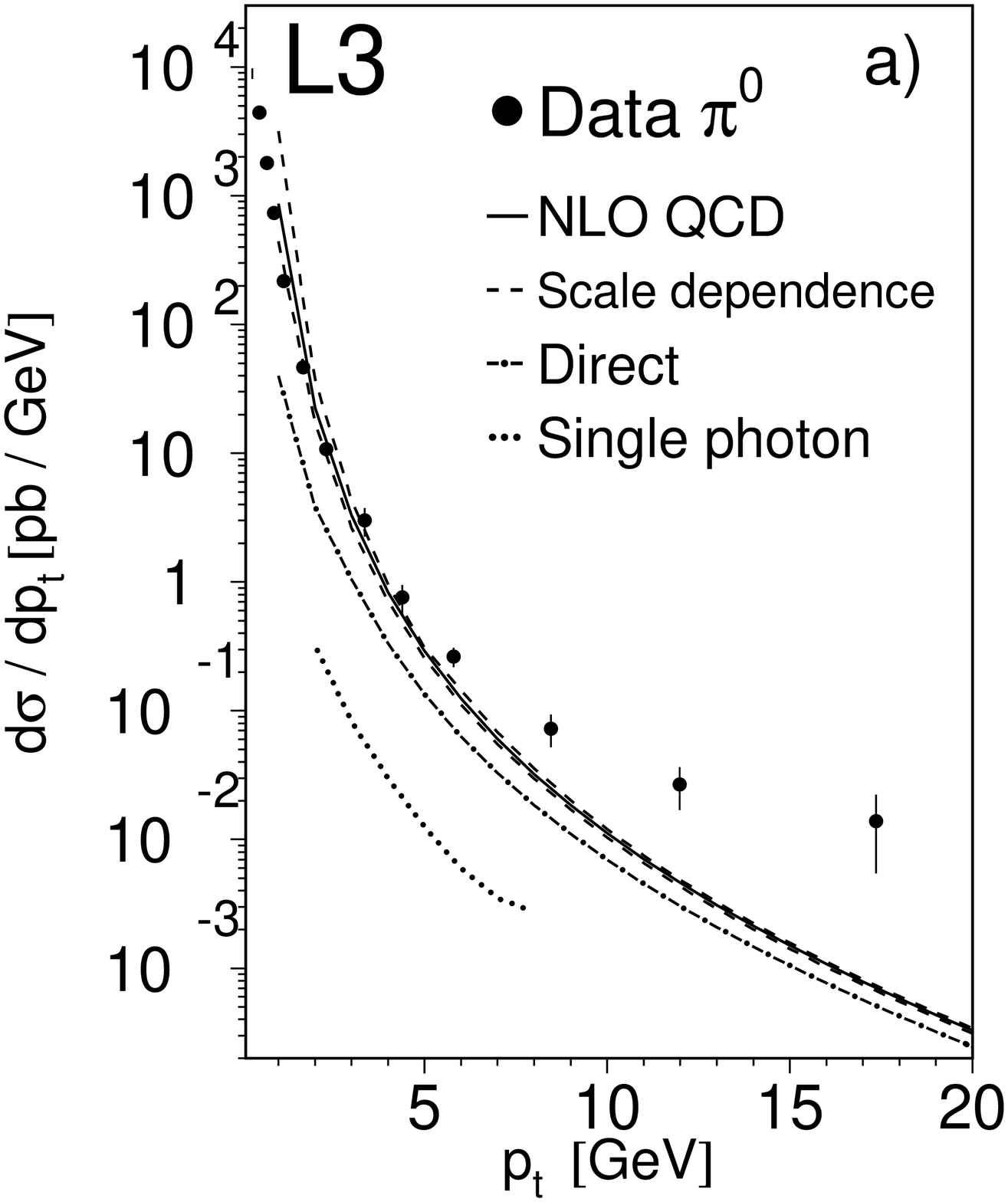,height=4.5cm}~\epsfig{file=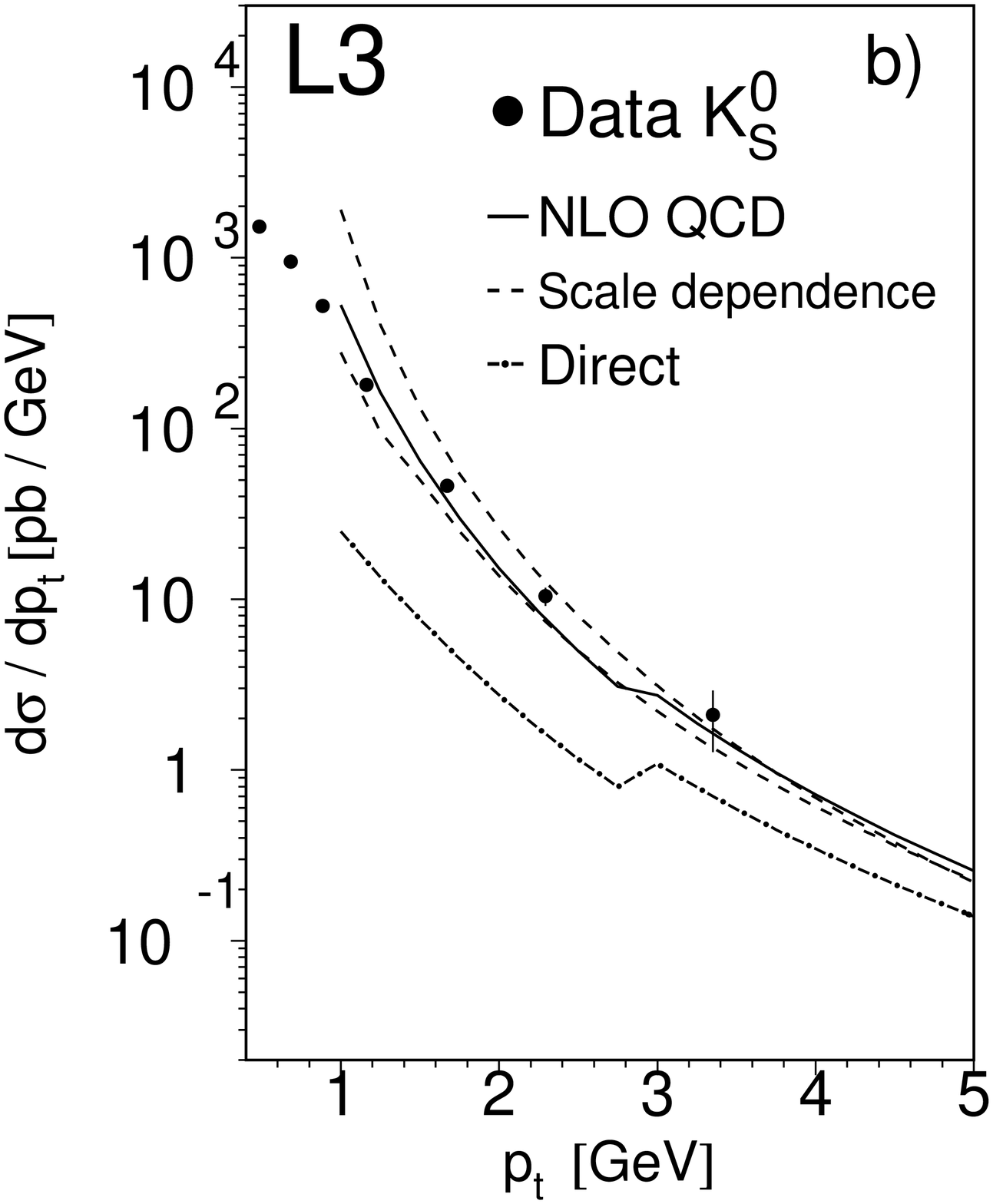,height=4.5cm}
\end{center}
\caption{Inclusive differential cross section, $d\sigma/dp_t$, compared to MC and NLO predictions 
and exponential and power law behaviour for $\pi^0$ and $K_s^0$ production.}
\label{fig:l3hadrons}
\end{figure}

Measurements of hadron production testing both the fragmentation~\cite{traynor} and production 
mechanism~\cite{achard,boogert} were presented. At LEP, the L3 collaboration have measured 
$\pi^0$ and $K_s^0$ production and compared them with predictions from NLO calculations, MC 
models and an exponential fall-off as shown in Fig.~\ref{fig:l3hadrons}. The cross sections 
as a function of the transverse momentum of the hadron, $p_t$, are described by an 
exponential at low $p_t$ and by a power law function, $Ap_t^{-B}$, at values larger than 1.5~GeV. 
The value of the power B is compatible with 4 for both $\pi^0$ and $K_s^0$ mesons as expected 
in $2 \rightarrow 2$ hard parton scattering~\cite{brodsky}. Both the MC and NLO predictions 
give qualitatively similar results. At low $p_t$ they describe the data well, but fail at 
high $p_t$ in the cross section for $\pi^0$ production. It would be of interest to see the 
distribution to higher values for the $K_s^0$ cross section, but this will be unattainable 
at LEP. 

At HERA large samples of $K_s^0$ mesons exist, extending up to transverse momenta of $\sim 8$~GeV, 
which could verify the preference of the power law also for this meson. The production of 
$K_s^0$ and $\Lambda$ hadrons has been considered at HERA and their relative rates 
compared to MC predictions. 
When the MC predictions are normalised to the measured $K_s^0$ production cross section, 
predictions from {\sc Herwig} overestimate and from {\sc Pythia} underestimate, the $\Lambda$ 
productions cross section.

%%%% \begin{figure}[htp]
%%%% \begin{center}
%%%% ~\epsfig{file=lam_hera.eps,height=5.5cm}
%%%% \end{center}
%%%% \caption{Cross section, $d\sigma/dx_gamma^{\rm obs}$, for $\Lambda$ production compared with 
%%%% {\sc Herwig} and {\sc Pythia} MC predictions. The MC is normalised both to the data and the 
%%%% cross section for $K_s^0$ production.}
%%%% \label{fig:zeushadrons}
%%%% \end{figure}

\subsection{Prompt photon production}

\begin{figure}[htp]
\begin{center}
~\epsfig{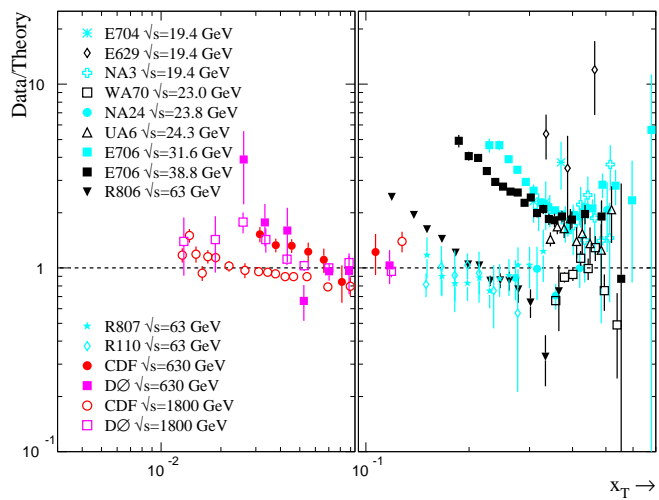}
\end{center}
\caption{Comparison of data and QCD predictions for prompt photon production in hadron-hadron 
collisions}
\label{fig:phot_had}
\end{figure}

The description of prompt photon production by QCD at hadron colliders has long been an issue 
of interest. The current status of comparison between data and theory is shown in 
Fig.~\ref{fig:phot_had} in which clear discrepancies are seen over a large range in 
energy~\cite{begel}. In particular, at the Tevatron, 
the data from both D0 and CDF is not well described, particularly at the lower transverse 
energies.
Explanations for this discrepancy in prompt photon production vary, for example the 
need for resummed calculations or extra $k_T$ in the colliding hadron being invoked. In 
this light, ZEUS have made an extraction of the $k_T$ in the proton using a sample of events 
with a prompt photon and an approximately back-to-back jet. This was extracted by considering 
variables sensitive to $k_T$  and fitting the {\sc Pythia} MC 
predictions to these distributions. The value found was consistent with a trend of increasing 
$k_T$ with increasing centre-of-mass energy. However, a recent NLO calculation~\cite{font} of 
prompt photon and jet production can describe the $k_T$-sensitive variables, with no extra 
$k_T$ required. In fact, calculations from the same group~\cite{aur} can describe most 
of the hadron-hadron measurements shown in Fig.~\ref{fig:phot_had}. Those data not described 
have found to be inconsistent with other measurements~\cite{aur}.
\begin{figure}[htp]
\begin{center}
~\epsfig{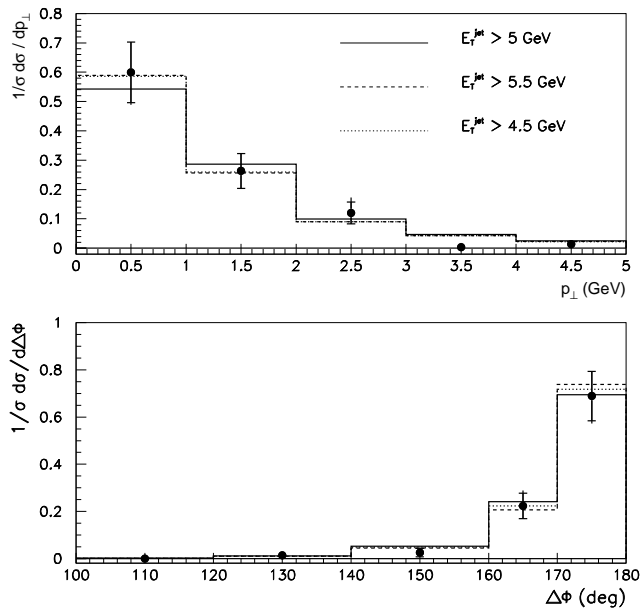}
\end{center}
\caption{Measured normalised cross sections for transverse momenta, $p_\perp$ and angle, 
$\Delta \phi$, of the photon relative to the jet compared to NLO.}
\label{fig:kt}
\end{figure}

\subsection{Jet production}

Jet production is being extensively studied in $\gamma \gamma$ collisions at LEP~\cite{gammagamma} 
and $ep$ collisions at HERA~\cite{herajets}. The measurements investigate the detailed 
dynamics of QCD, the parton densities in the photon and proton and allow measurements of the 
strong coupling constant, $\alpha_s$.

The dijet production cross section in $\gamma \gamma$ collisions is sensitive to the gluon 
content of the photon at LO. To try and constrain this, the cross section has been measured by 
both the DELPHI and OPAL collaborations~\cite{gammagamma}; the measurements are consistent with 
each other. 
\begin{figure}[htp]
\begin{center}
~\epsfig{file=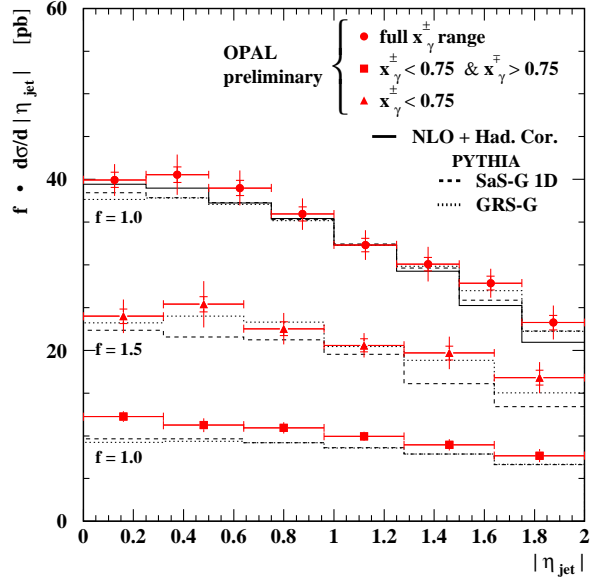,height=8cm}
\end{center}
\caption{Measured cross sections as a function of pseudorapidity, compared to predictions from 
NLO and {\sc Pythia} MC. The measurements are shown in different regions of $x_\gamma^\pm$.}
\label{fig:opal_jet}
\end{figure}
Using a luminosity of $\sim 600$ pb$^{-1}$, some results from OPAL are shown in 
Fig.~\ref{fig:opal_jet} compared to prediction from NLO and the {\sc Pythia} MC. In the region 
of high-$x_\gamma^\pm$, the NLO and {\sc Pythia} MC prediction describe the data well. 
For the case of single resolved events at low-$x_\gamma^\pm$, the {\sc Pythia} prediction 
underestimates the data by about $20\%$. The cross section in $x_\gamma$ is very sensitive 
to hadronisation effects and needs to be studied further. The accurate data can be used 
in future fits to constrain the parton and in particular the gluon density in the photon.

\begin{figure}[htp]
\begin{center}
~\epsfig{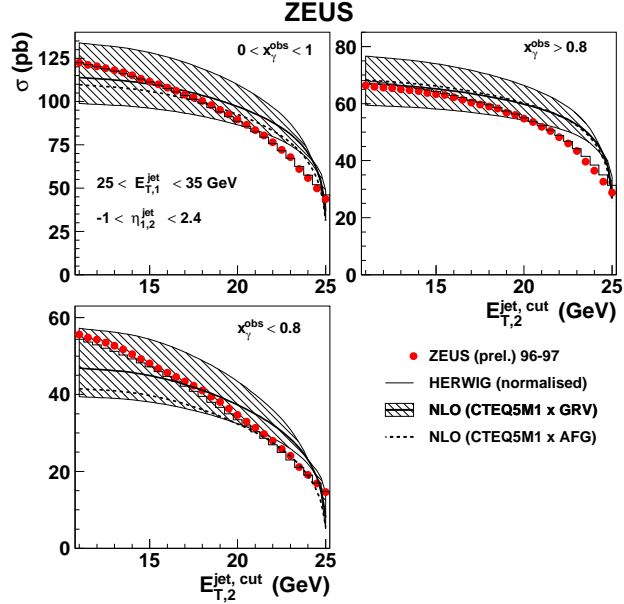}
\end{center}
\caption{Measured cross sections as a function of the cut on the second highest transverse energy 
jet, $E_{T,2}^{\rm jet, cut}$ compared to prediction from NLO and {\sc Herwig} MC.}
\label{fig:zeus_dijet}
\end{figure}

To make dijet measurements, the cuts on the jets have to be chosen carefully so as 
not to be infra-red sensitive. This has been extensively studied at HERA. The cuts on the 
two jets have to be sufficiently 
different to allow for soft-gluon emission. This region can be seen in Fig.~\ref{fig:zeus_dijet} 
at high values of $E_{T,2}^{\rm jet, cut}$, where the NLO prediction has very small 
uncertainties and overshoots the measured cross section. At low values of $E_{T,2}^{\rm jet, cut}$, 
the data and NLO have similar shapes for high-$x_\gamma^{\rm obs}$ and very different 
shapes for low-$x_\gamma^{\rm obs}$. Although the NLO and data agree within the uncertainties, 
the inclusion of higher orders would require a significant change in shape. This could also 
indicate problems with the photon parton densities, and higher order calculations would 
help further constrain them. It should be noted that the predictions from the {\sc Herwig} 
MC describe the shape of the data well. It is normalised to the total cross section for 
the full $x_\gamma^{\rm obs}$ region and $E_{T,2}^{\rm jet, cut} = $11 GeV.
\vspace{-0.4cm}
\section{Discussion}
\vspace{-0.1cm}
Some measurements of jet and hadron production have been discussed. Calculations to NLO 
QCD do not give a complete description of all the measurements. With the large data samples 
now collected at LEP and HERA, these precise measurements could be used to constrain parton 
density functions and tune MC's. Important theoretical work is needed in producing higher 
order calculations and/or NLO calculations with parton showers~\cite{poetter} and hadronisation. 
Both of these projects have started and are essential if theory is to keep up with the 
increasing precision of the data.

\end{document}